\documentclass[conference,twocolumn]{IEEEtran}

\newlength{\figwidth}
\usepackage{color}
\definecolor{links}{rgb}{0.7,0,0}   
\definecolor{urls}{rgb}{0,0,0.8}    
\definecolor{cites}{rgb}{0,0,0.8}   

\usepackage{cite}        
\usepackage{url} 
\usepackage[intlimits]{amsmath}
\usepackage{bbm}
\usepackage{graphicx}
\usepackage{paralist}
\usepackage{fancyref}
\usepackage[stretch=16,shrink=16,step=4]{microtype}
\usepackage{vmr-symbols-vecbold}
\usepackage{standard-macros}
\usepackage{steinmetz}
\usepackage{psfrag}
\usepackage{flushend}

\usepackage[colorlinks,hyperindex,linkcolor=links,citecolor=cites,urlcolor=urls]{hyperref} 

\makeatletter
\def\@IEEEinterspaceratioM{0.265}
\def\@IEEEinterspaceMINratioM{0.1651}
\def\@IEEEinterspaceMAXratioM{0.38}

\def\@IEEEinterspaceratioB{0.31}
\def\@IEEEinterspaceMINratioB{0.19}
\def\@IEEEinterspaceMAXratioB{0.38}
\@IEEEtunefonts
\makeatother
\newcommand{\ES}{\mathrm{E}}

\newcommand{\parl}{\left(}
\newcommand{\parr}{\right)}

\begin{document}

\IEEEoverridecommandlockouts

\title{\huge{Receiver Algorithm based on Differential Signaling \\for SIMO Phase Noise Channels \\with Common and Separate Oscillator Configurations}}
%
%
\author{M.~Reza~Khanzadi$^{\textrm{\dag,*}}$, Rajet~Krishnan$^{\textrm{*}}$, Thomas~Eriksson$^{\textrm{*}}$
\\	\IEEEauthorblockA{
\textdagger{Department of Microtechnology and Nanoscience, Microwave Electronics Laboratory}\\
\textasteriskcentered{Department of Signals and Systems, Communication Systems Group}
\\{Chalmers University of Technology, Gothenburg, Sweden}
\\{\{khanzadi, rajet, thomase\}@chalmers.se}
}
}

%
%
%
%
\maketitle

\begin{abstract}
In this paper, a receiver algorithm consisting of differential transmission and a two-stage detection for a single-input multiple-output (SIMO) phase-noise channels is studied. Specifically, the phases of the QAM modulated data symbols are manipulated before transmission in order to make them more immune to the random rotational effects of phase noise. At the receiver, a two-stage detector is implemented, which first detects the amplitude of the transmitted symbols from a nonlinear combination of the received signal amplitudes. Then in the second stage, the detector performs phase detection. The studied signaling method does not require transmission of any known symbols that act as pilots. Furthermore, no phase noise estimator (or a tracker) is needed at the receiver to compensate the effect of phase noise. This considerably reduces the complexity of the receiver structure. Moreover, it is observed that the studied algorithm can be used for the setups where a common local oscillator or separate independent oscillators drive the radio-frequency circuitries connected to each antenna. Due to the differential encoding/decoding of the phase, weighted averaging can be employed at a multi-antenna receiver, allowing for phase noise suppression to leverage the large number of antennas. Hence, we observe that the performance improves by increasing the number of antennas, especially in the separate oscillator case. Further increasing the number of receive antennas results in a performance error floor, which is a function of the quality of the oscillator at the transmitter.
\end{abstract}
\section{Introduction}
\label{sec:intro}
Phase noise due to phase and frequency instability in the local radio-frequency (RF) oscillators used in wireless communication links results in synchronization issues, which degrade the system performance \cite{Colavolpe2005,Khanzadi2013_1}. These effects are more severe when high-order modulation schemes are used in order to attain high spectral efficiency \cite{Krishnan2013_1}. It is also known that phase noise in RF oscillators increases with frequency~\cite{Mehrpouyan2014_EBAND,khanzadi2014_highFreq}.

In multiple-antenna systems, the impact of phase noise is different depending on whether the RF circuitries connected to each antenna are driven by separate (independent) local oscillators (SLO) or by a common local oscillator (CLO). Although the CLO configuration is of lower implementation complexity, the SLO configuration is unavoidable when a spacing, as large as a few meters, is needed between the antennas in order to exploit the available spatial degrees of freedom for higher multiplexing or diversity gains \cite{gesbert02-12a,chizhik-02-01a,durisi13-02a,krishnan2013_MIMO_algorithms,khanzadi2014_SLO_CLO}.

Carrier phase synchronization in multiple-input multiple-output (MIMO) systems has been extensively studied in the literature (e.g., \cite{ Schenk2005,Liu2006,Pedersen2008,Krishnan2012_1, Mehrpouyan2012, Khanzadi2013_0_COMP, krishnan2014_SP_letter,krishnan2015_VT_linear_MIMO} and the references therein). Phase synchronization methods in the case of the CLO configuration is similar to that in the case of a single-antenna system\cite{Colavolpe2005,Khanzadi2013_1,Krishnan2013_1}, which is followed by data detection that is generally used for MIMO systems \cite{larsson2009mimo_detection}. 
However, designing receiver algorithms for joint phase noise estimation and data detection in the SLO case is more challenging. This is mainly due to the fact that first, coherent combining of the received signals at the receiver is not possible. Next, multiple phase noise parameters must be tracked and compensated before performing data detection. The effect of phase noise on the performance of MIMO systems with a SLO configuration is studied in \cite{Krishnan2012_1,krishnan2013_MIMO_algorithms, krishnan2015_VT_linear_MIMO}.  Authors in \cite{Mehrpouyan2012} investigate bounds on the mean-square error performance of phase noise estimators, including the extended Kalman filter (EKF).  In~\cite{krishnan2013_MIMO_algorithms}, a joint phase noise estimation, data detection method based on maximum a posteriori (MAP) theory is proposed by applying the sum-product algorithm. However, most of the techniques proposed in prior studies suffer from high complexity, and are based on pilot-based transmissions for estimation and detection. 

In this work, we consider the scenario where a single-antenna user communicates on the uplink with a base station with multiple antennas over an AWGN channel impaired by phase noise. For this scenario, we develop an algorithm that consists of differential transmission, followed by a two-stage amplitude and phase detector for detection of the transmitted signal~(see~\cite{Krishnan2013_1} for two-stage detection). Differential transmission can be performed without the aid of pilot symbols, and no additional phase noise estimator/trackers are required at the receiver in order to track the random phase noise. Moreover, it has lower complexity than schemes, which are based on the MAP theory. More importantly, due to the differential encoding/decoding of transmitted phase, weighted averaging can be employed at a multi-antenna receiver, allowing for phase noise suppression to leverage the large number of antennas. These aspects make the developed algorithm particularly attractive for systems~i)~that require low-complexity receiver algorithms,~ii)~whose performance are affected by the use of pilots, as in massive-antenna systems~\cite{chowdhury2014design,ErikLarsson_CommMag_2014_a1}. We observe that the developed method is invariant to the oscillator configuration at the base station when applied to a SIMO phase noise channel. 
Our simulation results show that this method outperforms a Kalman filter-based algorithm~\cite{Mehrpouyan2012} at high SNR or high phase noise scenarios. We also observe that when the SNR is high, the SLO yields a lower symbol error rate than the CLO because of the noise averaging effects. Furthermore, the performance of the studied method in SLO configuration improves by increasing the number of antennas till it achieves an error floor that is a function of the level of phase noise at the transmitter. This observation is in line with the results from prior work on the performance of uplink transmissions in presence of phase noise (see e.g.,\cite{bjornson13-07a,krishnan2014_SP_letter,khanzadi2014_SLO_CLO}).

\emph{Notations:}
Italic letters $(x)$, boldface letters $(\mathbf{ x})$, and uppercase boldface letters $(\mathbf{X})$ denote scalar variables, vectors, and matrices, respectively. The $(i,j)$th entry of matrix $(\mathbf{X})$ is denoted by $[\mathbf{X}]_{i,j}$.
With $\normal(0,\sigma^2)$ and $\jpg(0,\sigma^2)$, we denote the probability distribution of a real Gaussian, and of a circularly symmetric complex Gaussian random variable with zero mean and variance $\sigma^2$.
For a given complex vector $\vecx$, we  denote by $\phase{\vecx}$, the vector that contains the phase of the elements of $\vecx$; $\Re(\cdot)$, and $\Im(\cdot)$, denote the real and imaginary parts of complex values, respectively; $\vecnorm{\cdot}$ denotes the Euclidean norm of a vector; the operator $\diag{(\cdot)}$ generates a diagonal matrix from a vector; $I_\nu$ denotes the modified Bessel functions of the first kind with order $\nu$; $Q(\cdot)$ is the Q-function.

\section{System Model}
\label{sec:system_model}
Consider the uplink channel of a wireless communication system where a single-antenna user communicates with a base station equipped with $M$ antennas over an AWGN channel impaired by phase noise. This yields the following $1 \times M$ single-input multiple-output (SIMO) phase-noise channel:
\begin{IEEEeqnarray}{rCL}\label{eq:simo_io}
  \vecy_k=\bTheta_k\vech x_k+\vecw_k, \quad k=1,\dots,n,
\end{IEEEeqnarray}
where $x_k$ is the transmitted symbol in the $k$th time instant,  $\bTheta_k=\diag{([e^{\jmath\theta_{1,k}},\dots,e^{\jmath\theta_{M,k}}])}$, and $\theta_{m,k}=\theta^{[\mathrm{r}]}_{k}+\theta^{[\mathrm{t}]}_{k}$, for  $m=1,\dots,M$, where $\{\theta^{[\mathrm{t}]}_{k}\}$ and $\{\theta^{[\mathrm{r}]}_{k}\}$ represent the random phase noise samples, in the $k$th  time instant, from the local oscillators at the transmitter and receiver, respectively. We model the phase noise as a Wiener process \cite{Demir2006}
\begin{IEEEeqnarray}{rCL}
&&\theta^{[\mathrm{t}]}_{k}=\theta^{[\mathrm{t}]}_{k-1}+\Delta^{[\mathrm{t}]}_{k-1}\label{eq:Wiener_process_t}\\
&&\theta^{[\mathrm{r}]}_{k}=\theta^{[\mathrm{r}]}_{m,k-1}+\Delta^{[\mathrm{r}]}_{k-1},\label{eq:Wiener_process_r}
\end{IEEEeqnarray}
where $\{\Delta^{[\mathrm{t}]}_k\}$ and $\{\Delta^{[\mathrm{r}]}_k\}$ are the Gaussian random samples drawn independently from~$\normal (0,\sigma^2_{\Delta^{[t]}})$ and~$\normal (0,\sigma^2_{\Delta^{[r]}})$ distributions, respectively\footnote{For discussions on the limitations of this model see \cite{Khanzadi2011,Khanzadi2013_2_ColoredPNEst,Khanzadi2013_1} and the references therein.}.
The vector~$\vech=\tp{[h_1,\dots, h_{M}]}$ contains the path-loss coefficients, which are assumed to be deterministic, time-invariant, and known to the receiver. Finally, the entries of $\vecw_k=\tp{[w_{1,k},\dots, w_{M,k}]}$ are the AWGN samples, which are drawn independently from a $\jpg(0,2)$ distribution.\footnote{As we shall see, normalizing the noise variance to $2$ will turn out convenient.}

\section{SLO Configuration}
\label{sec:SLO_Conf}
In the SLO configuration, coherent combining of the received signals is not possible. 
In this section, we study a differential transmission algorithm, which is followed by a two-stage data detection scheme at the receiver. 
The studied signaling method can be used along with the various modulation schemes such as $M$-QAM. This method enables a non-data-aided carrier phase synchronization at the receiver over the SIMO phase-noise channel.
\begin{figure*}
\centering
\psfrag{sk}[cc][][1]{$s_k$}%
\psfrag{xk}[cc][][1]{$x_k$}%
\psfrag{pk}[cc][][1]{$\phi_k=\phase{s_k}$}%
\psfrag{ak}[cc][][1]{$r_k=\abs{s_k}$}%
\psfrag{z}[cc][][1]{$z^{-1}$}%
\psfrag{exp}[cc][][1]{$e^{\jmath(\cdot)}$}%
\psfrag{pntk}[cc][][1]{$e^{\jmath\theta^{[\mathrm{t}]}_k}$}%
\psfrag{pnr1k}[cc][][1]{$e^{\jmath\theta^{[\mathrm{r}]}_{1,k}}$}%
\psfrag{pnrmk}[cc][][1]{$e^{\jmath\theta^{[\mathrm{r}]}_{M,k}}$}%
\psfrag{w1}[cc][][1]{$w_{1,k}$}%
\psfrag{wm}[cc][][1]{$w_{M,k}$}%
\psfrag{y1}[cc][][1]{$y_{1,k}$}%
\psfrag{ym}[cc][][1]{$y_{M,k}$}%
\psfrag{py1}[cc][][1]{$\phase{y_{1,k}}$}%
\psfrag{pym}[cc][][1]{$\phase{y_{M,k}}$}%
\psfrag{py1}[cc][][1]{$\phase{y_{1,k}}$}%
\psfrag{pym}[cc][][1]{$\phase{y_{M,k}}$}%
\psfrag{ay1}[cc][][0.85]{$\abs{y_{1,k}}^2$}%
\psfrag{aym}[cc][][0.85]{$\abs{y_{M,k}}^2$}%
\psfrag{tk}[cc][][1]{$t_k$}%
\psfrag{rhk}[cc][][1]{$\hat{r}_k$}%
\psfrag{phk}[cc][][1]{$\hat{\phi}_k$}%
\psfrag{shk}[cc][][1]{$\hat{s}_k$}%
\psfrag{h1}[cc][][1]{$h_1$}%
\psfrag{hm}[cc][][1]{$h_M$}%
\includegraphics [height=3.1in]{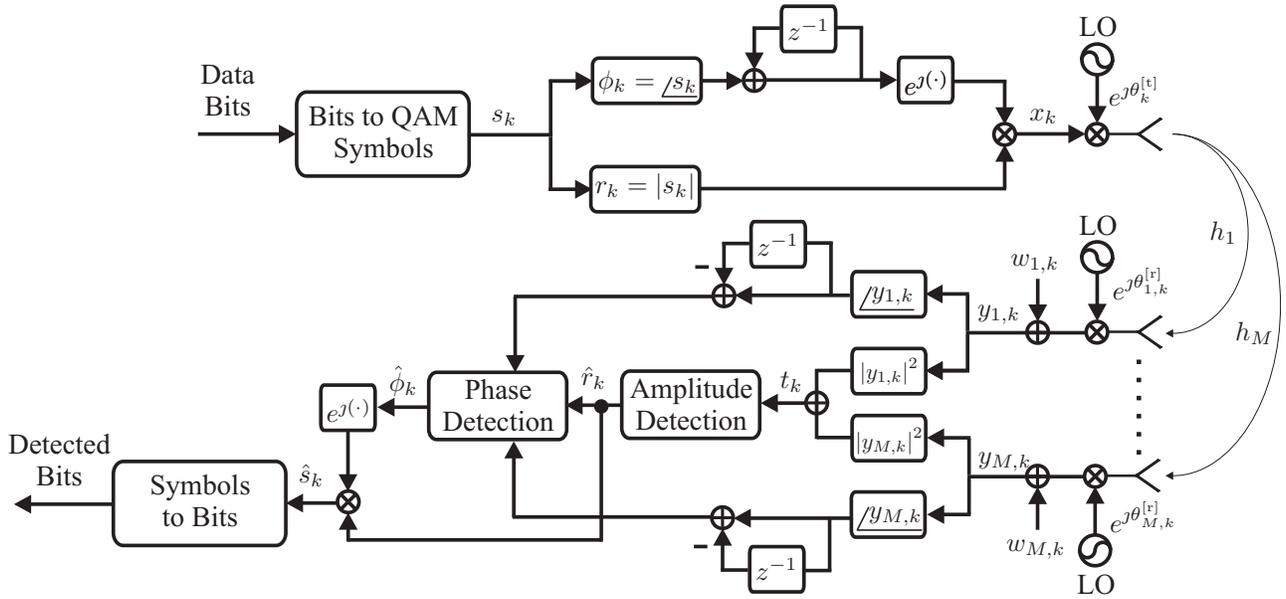}
\caption{System diagram of the proposed differential signaling method for an $1 \times M$ SIMO channel.}
\label{fig:sys_model}
\end{figure*}

As illustrated in Fig.~\ref{fig:sys_model}, at the transmitter side, we first modulate the data bits as QAM symbols denoted as~$s_k$, which in the polar coordinate can be expressed as~$s_k=r_k e^{\jmath \phi_k}$. Next, we manipulate the phase of each symbol before sending it over the SIMO phase noise channel. More specifically, the phase of the transmitted symbol is generated by accumulating the phase of the current symbol and that of the previously transmitted symbol, i.e., 
\begin{IEEEeqnarray}{rCL}\label{eq:tx_phase_accumulation}
	x_k=r_k e^{\jmath\sumo{l}{k}\phi_k}.
\end{IEEEeqnarray}
Note that the amplitudes of the symbols remain unaltered. 

At the receiver side, we propose a two-stage detection procedure in order to detect the transmitted QAM symbols. First, a symbol-by-symbol maximum a posteriori (MAP) detection of the amplitude is performed. Then the transmitted signal amplitude, as determined by the amplitude detector, is used along with the phase of the received signals at each antenna in order to perform differential maximum likelihood (ML) phase detection.
It is straightforward to show that given~$r_k$, ~$t_k=\vecnorm{\vecy_k}^2$ follows a noncentral chi-squared distribution with $2M$ degrees of freedom and noncentrality parameter~$r_k^2\vecnorm{\vech}^2$ \cite{khanzadi2014_SLO_CLO}, i.e.,
\begin{IEEEeqnarray}{rCL}\label{eq:chi_squared_variable}
{t_k}&\distas&\big|r_k\vecnorm{\vech}+w_1\big|^2+\sum_{m=2}^{M}\abs{w_m}^2,
\end{IEEEeqnarray}
where $\distas$ denotes equality in distribution, and $\{w_m\}$, $m=1,\dots,M$ are independently and identically distributed (\iid) from $\jpg(0,2)$. The MAP amplitude detector is determined as
\begin{IEEEeqnarray}{rCL}\label{eq:map_amp_detector}
	\hat{r}_k &=&\arg \underset{r_k}{\max}~f(r_k\given t_k)=\arg \underset{r_k}{\max}~f(t_k\given r_k)p(r_k),
\end{IEEEeqnarray}
where $p(r_k)$ denotes the probability mass function of $r_k$.
The likelihood function $f(t_k\given r_k)$ can be determined from the definition of a noncentral chi-squared distribution~\cite[Eq.~2.44]{simon2007probability}
\begin{IEEEeqnarray}{rCL}\label{eq:chi_dist_def}
f(t_k\given r_k)=&&\notag\\
&&\hspace{-1.2cm}\frac{1}{2}e^{-(t_k+r_k^2\vecnorm{\vech}^2)}\left(\frac{t_k}{r_k^2\vecnorm{\vech}^2}\right)^{(M-1)/2} \hspace{-0.6cm}I_{M-1}\left(r_k \vecnorm{\vech}\sqrt{t_k}\right)\hspace{-0.1cm}.
\end{IEEEeqnarray}
The detector can be implemented by substituting \eqref{eq:chi_dist_def} in \eqref{eq:map_amp_detector}, and then finding the maximizer by evaluating \eqref{eq:map_amp_detector} over all possible symbol amplitudes. For example, there are three amplitude possibilities when using $16$-QAM,
\begin{IEEEeqnarray}{rCL}\label{eq:rk_possiblities}
&&r_k\in\left\{\sqrt{\frac{\ES}{5}},\sqrt{\ES},3\sqrt{\frac{\ES}{5}}\right\},
\end{IEEEeqnarray}
with
\begin{IEEEeqnarray}{rCL}
&&p(r_k=\sqrt{\frac{\ES}{5}})=p\parl r_k=3\sqrt{\frac{\ES}{5}} \parr =\frac{1}{2}p\parl r_k=\sqrt{\ES}\parr, \IEEEeqnarraynumspace
\end{IEEEeqnarray}
where $\ES$ denotes the average symbol energy.

In the next stage, a differential phase detection is performed as follows. The phase of the received signal per each antenna is written as \cite[Eq.~12]{Krishnan2012_1}
\begin{IEEEeqnarray}{rCL}
&&\hspace{-0.5cm}\phase{y_{m,k}}\notag\\
&=&\phase{x_k}+\theta_{m,k}+\phase{h_m}+\arctan\frac{\Im\{w_{m,k}\}}{r_k\abs{h_m}+\Re\{w_{m,k}\}}\\
&\approx&\phase{x_k}+\theta_{m,k}+\phase{h_m}+\tilde{w}_{m,k},\label{eq:y_phase_approx}
\end{IEEEeqnarray}
where~$\tilde{w}_{m,k}=\Im\{w_{m,k}\}/(r_k\abs{h_m})$ is a zero-mean real Gaussian random variable. The approximation in \eqref{eq:y_phase_approx} is valid at moderate and high SNR~\cite{Krishnan2012_1}.
In the first step of phase detection, from the phase of the current received signal, we subtract the phase of the previously received signal at each receive antenna as
\begin{IEEEeqnarray}{rCL}
\hspace{-.7cm}\phase{y_{m,k}}-\phase{y_{m,k-1}}&\approx&\phase{x_k}-\phase{x_{k-1}}+\theta_{m,k}-\theta_{m,k-1}\notag\\&&+\tilde{w}_{m,k}-\tilde{w}_{m,k-1}\\
&=&\phi_k+\Delta^{[\mathrm{t}]}_{k-1}+\Delta^{[\mathrm{r}]}_{k-1}\notag\\&&+\tilde{w}_{m,k}-\tilde{w}_{m,k-1},\label{eq:phase_diff}
\end{IEEEeqnarray}
where the equality in the last step is obtained by using \eqref{eq:y_phase_approx}, \eqref{eq:tx_phase_accumulation}, \eqref{eq:Wiener_process_t}, and \eqref{eq:Wiener_process_r}. By means of this differential transformation, we mitigate the effect of instantaneous unknown phase noise values on the phase of the transmitted symbols. From \eqref{eq:phase_diff} we observe that~$\phase{y_{m,k}}-\phase{y_{m,k-1}}$, for $m=1,\dots,M$ are independent, and approximately follow a Gaussian distribution. Hence, the sufficient statistic for detection of~$\phi_k$ is given by~\cite{book_kay_est}
\begin{IEEEeqnarray}{rCL}\label{eq:phi_sufficient_stat}
\psi_k&=&\frac{1}{M}\sumo{m}{M}\Big(\phase{y_{m,k}}-\phase{y_{m,k-1}}\Big).
\end{IEEEeqnarray}
It is straightforward to show from~\eqref{eq:phase_diff} and~\eqref{eq:phi_sufficient_stat} that
\begin{IEEEeqnarray}{rCL}\label{eq:pdf_phi_suff_stat}
f(\psi_k|\phi_k)&=&\normal (\phi_k,\sigma^2_{\psi_k}),
\end{IEEEeqnarray}
where 
\begin{align}\label{eq:pdf_phi_suff_stat_variance}
&\sigma^2_{\psi_k}=\notag\\
&\begin{cases} \sigma^2_{\Delta^{[t]}} + \sigma^2_{\Delta^{[r]}}+ \frac{1}{M^2}(\sumo{m}{M}\frac{1}{|h_m|^2})(\frac{1}{r_k^2}+\frac{1}{r_{k-1}^2})\quad\quad\mbox{CLO}\\
\sigma^2_{\Delta^{[t]}} + \frac{1}{M}\sigma^2_{\Delta^{[r]}}+ \frac{1}{M^2}(\sumo{m}{M}\frac{1}{|h_m|^2})(\frac{1}{r_k^2}+\frac{1}{r_{k-1}^2}) \quad\mbox{SLO}\end{cases}\hspace{-0.4cm}. 
\end{align}
As we see in~\eqref{eq:pdf_phi_suff_stat_variance}, in the SLO case, the total variance of phase noise from receiver's local oscillators reduces upon increasing $M$. This is because the receiver phase-noise samples are \iid over receive antennas.~\footnote{Note that \eqref{eq:phi_sufficient_stat} is an equally-weighted sum of phases over the receive antennas. However, the weights can be optimized in order to deal with differences in quality of the signals received at the different antennas. Similarly, the amplitude detection can be improved by considering an optimized weighted sum of amplitude squares of received signals.}  

In order to detect $\phi_k$ we perform a maximum likelihood detection as follows:
\begin{IEEEeqnarray}{rCL}
\hat{\phi}_k&=&\underset{\phi_k\in \phase{\mathcal{C}({\hat{r}_k)}}}{\arg \max}f(\psi_k\given \phi_k)\\
&=&\underset{\phi_k\in \phase{\mathcal{C}({\hat{r}_k})}}{\arg \min}(\psi_k-\phi_k)^2,\label{eq:Phase_Euqlidian_distance}
\end{IEEEeqnarray}
where the minimization in \eqref{eq:Phase_Euqlidian_distance} is performed over the phase of a set of transmitted QAM symbols with amplitude $\hat{r}_k$ denoted as $\phase{\mathcal{C}({\hat{r}_k})}$. Note that the phase detection in \eqref{eq:Phase_Euqlidian_distance} can be seen as a Euclidean distance detector in the phase domain.
\section{CLO Configuration}
\label{sec:CLO_Conf}
In the CLO configuration, we have that~$\theta_{1,k}=\dots=\theta_{M,k}=\theta_{k}$ for all $k$. Therefore, the input-output relation \eqref{eq:simo_io} simplifies to
\begin{IEEEeqnarray}{rCL}\label{eq:simo_common_osc_io}
  \vecy_k=e^{\jmath\theta_k}\vech x_k+\vecw_k.
\end{IEEEeqnarray}
As the $M$ signals received in a given time instant are affected by only one phase noise process, it is possible to first coherently combine the received signals by projecting~$\vecy_k$ on~$\vech/\vecnorm{\vech}$, and convert \eqref{eq:simo_common_osc_io} to a single-input single-output (SISO) channel, followed by the compensation of $\theta_k$ and data detection.
Note that the CLO case, after coherent combining, is a special case of the SLO configuration with $M=1$. 
\section{Symbol Error Probability Analysis}
\label{sec:serAnalysis}
In this section, we derive analytical expressions for the symbol error probability (SEP) performance of the developed differential algorithm and the two-stage detector for arbitrary constellations. Consider the transmission of an arbitrary constellation with $N$ equally-likely symbol points in~\eqref{eq:simo_io}. The symbol error probability $P_\text{e}$ for this constellation is upper-bounded by averaging over all pair-wise symbol error probabilities (union bound)~\cite{Proakis2008} as
\begin{IEEEeqnarray}{rCL}\label{eq:Pe_union_bound}
P_\text{e}&=&\frac{1}{N}\sumo{i}{N}\sum_{j=1,j\neq i}^{N}P(E_{ij}),
\end{IEEEeqnarray}
where~$P(E_{ij})$ is the pair-wise symbol error probability. This corresponds to the probability of the error event $E_{ij}$, where the symbol $j$ is detected given that symbol~$i$ has been transmitted. For a two-stage detector, the pair-wise symbol error is a result of two events: the error in pair-wise detection of the amplitudes, denoted as~$E_{ij}^\text{amplitude}$, and the error in detection of the phase of the transmitted symbol, denoted as~$E_{ij}^\text{phase}$. Accordingly, the pair-wise symbol error probability becomes
\begin{IEEEeqnarray}{rCL}\label{eq:Pe_pairwise_amp_phase}
P(E_{ij})=P(E_{ij}^\text{amplitude})P(E_{ij}^\text{phase}).
\end{IEEEeqnarray}

In order to simplify the analysis, we consider a high-SNR scenario, where no amplitude detection error occurs, i.e., $P(E_{ij}^\text{amplitude})=0$. In this case, the error in symbol detection occurs only in the phase domain. Therefore, $P(E_{ij}^\text{amplitude})=0$ for symbols $i$ and $j$ with different amplitudes, while $P(E_{ij}^\text{amplitude})=1$ for equal-amplitude symbols. By using~\eqref{eq:pdf_phi_suff_stat}, the pair-wise phase detection error probability for equal-amplitude symbols can be found as 
\begin{IEEEeqnarray}{rCL}\label{eq:Pe_pairwise_phase}
P(E_{ij}^\text{phase})=Q\lefto(\frac{|\phi_k^i-\phi_k^j|}{2\sigma_{\psi_k}}\right),
\end{IEEEeqnarray}
where~$\phi_k^i$ and~$\phi_k^j$ denote the phase of symbols $i$ and $j$, respectively. Note that the value of~$\sigma_{\psi_k}$, which is computed from~\eqref{eq:pdf_phi_suff_stat_variance} depends on amplitude of the current symbol and that of the previously transmitted symbol. In order to simplify the analysis and remove the dependency on the past symbol, we compute~$\sigma_{\psi_k}$ by setting~$r_{k-1}=\ES$ (the average symbol energy).

\paragraph*{Error floor at high-SNR, when $M\to \infty$}
In order to determine the SEP at high-SNR for large number of antennas, the pair-wise probability of error can be simplified by evaluating 
\begin{IEEEeqnarray}{rCL}\label{eq:Pe_pairwise_phase_large_M}
\lim_{M\to\infty} P(E_{ij}^\text{phase})=Q\lefto(\frac{|\phi_k^i-\phi_k^j|}{2\check{\sigma}_{\psi_k}}\right),
\end{IEEEeqnarray}
where from~\eqref{eq:pdf_phi_suff_stat_variance}
\begin{align}\label{eq:pdf_phi_suff_stat_variance_large_M}
\check{\sigma}^2_{\psi_k}=\lim_{M\to\infty}{\sigma}^2_{\psi_k} =\begin{cases} \sigma^2_{\Delta^{[t]}} + \sigma^2_{\Delta^{[r]}}&\quad\quad\mbox{CLO}\\
\sigma^2_{\Delta^{[t]}}&\quad\quad\mbox{SLO}\end{cases}. 
\end{align}
We can observe from~\eqref{eq:pdf_phi_suff_stat_variance_large_M} that the error floor in the CLO case is a function of phase noise innovation variance in transmitter and receiver oscillators, while  in the case of SLO configuration, it is only a function of phase noise innovation variance of transmitter oscillator.
\section{Overview of the Kalman Filter Method}
\label{sec:Kalman_method}
In this section we present a EKF-based phase noise estimation-data detection algorithm similar to that of developed in~\cite{Mehrpouyan2012}. The QAM symbols are transmitted directly over the phase-noise channel \eqref{eq:simo_io} while an EKF at the receiver compensates the effect of phase noise before data detection. Note that the QAM symbols transmitted here have not been manipulated as in \eqref{eq:tx_phase_accumulation}.

As already mentioned, in the SLO configuration, coherent combining of the received signals is not possible, and the vector $\vectheta_k=\tp{[\theta_{1,k},\dots,\theta_{M,k}]}$ consisting of $M$ phase noise processes must be tracked. The state equation in this case is
\begin{IEEEeqnarray}{rCL}\label{eq:vector_state_equation}
  \vectheta_k=\vectheta_{k-1}+\boldsymbol{\Delta}_{k-1},
\end{IEEEeqnarray}
where $\boldsymbol{\Delta_k}\sim\normal (\mathbf{0},\boldsymbol{\Sigma})$ and
\begin{IEEEeqnarray}{rCL}\label{eq:vector_state_sigma}
  [\boldsymbol{\Sigma}]_{i,j}=
\begin{cases} \sigma^2_{\Delta^{[t]}}+\sigma^2_{\Delta^{[r]}}~&\mbox{if } i=j\\
\sigma^2_{\Delta^{[t]}}~&\mbox{if } i\neq j
\end{cases},
\end{IEEEeqnarray}
for~$i,j\in\{1,\dots,M\}.$
%
We use \eqref{eq:vector_state_sigma} along with the observation model \eqref{eq:simo_io} to estimate $\vectheta_k$ by using an EKF~\cite{book_kay_est}. Note  that when pilot symbols are not transmitted, the EKF does not have reliable estimates of $x_k$, thereby causing the performance of the EKF to deteriorate. In order to avoid this problem, we transmit pilot symbols, which are known at the receiver. More specifically, at each time instant, we first perform the prediction step of the EKF which is independent of the observation. If a pilot symbol is transmitted, we use the pilot transmission to perform the update step. For the data symbols, we first use the predicted phases to de-rotate the received signal at each antenna. Then a Euclidean-distance based detector is used to detect the transmitted symbol~\cite{Krishnan2012_1}. Finally, the detected symbol is used in the update step of the EKF.
\section{Results and Discussions}
\label{sec:results}
In this section, we present the performance results of the proposed signaling method using $16$-QAM in terms of SEP versus SNR per symbol~($\ES/2$). We scale down the transmit power by $M$ when increasing the number of receive antennas to keep the average receive SNR constant. In our simulations we consider a quasi-static fading model~\cite[p.~2631]{biglieri98-10a},~\cite[Sec.~5.4.1]{tse05a} for the channel, where the~$\{h_m\}$ are independently drawn from a~$\jpg(0,1)$ distribution. The channel is kept a constant over each simulation trial, and is known to the receiver. The variance of the phase noise innovation for transmitter and receiver oscillators are set equal, unless it is explicitly mentioned.

\begin{figure}[t]
\centering
\psfrag{serl}[cc][][1]{SEP}%
\psfrag{snrl}[cc][][1]{SNR per Symbol~[dB]}%
\psfrag{EKF1}[cc][][0.7][-5]{EKF~$\sigma^2_\Delta=0.01$[rad$^2$]}%
\psfrag{EKF2}[cc][][0.7][-17]{EKF~$\sigma^2_\Delta=0.005$[rad$^2$]}%
\psfrag{DIF1}[cc][][0.7][-15]{DIF~$\sigma^2_\Delta=0.01$[rad$^2$]}%
\psfrag{DIF2}[cc][][0.7][-25]{DIF~$\sigma^2_\Delta=0.005$[rad$^2$]}%
\psfrag{EKF3}[cc][][0.7]{EKF~$\sigma^2_\Delta=0.001$[rad$^2$]}%
\psfrag{DIF3}[cc][][0.7]{DIF~$\sigma^2_\Delta=0.001$[rad$^2$]}%
\includegraphics [width=3.5in]{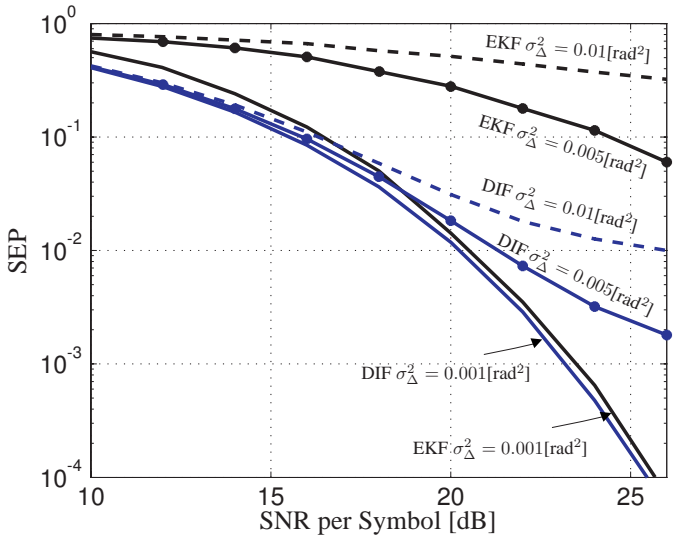}
\caption{SEP versus SNR for a $1\times 10$ SIMO channel with the CLO configuration.}
\label{fig:EKF_vs_DIFF}
\end{figure}
\begin{figure}[t]
\centering
\psfrag{serl}[cc][][1]{SEP}%
\psfrag{snrl}[cc][][1]{SNR per Symbol~[dB]}%
\psfrag{EKF1}[cc][][0.7]{EKF}%
\psfrag{DIF1}[cc][][0.7]{DIF}%
\psfrag{CLO}[cc][][0.7]{CLO}%
\psfrag{SLO}[cc][][0.7]{SLO}%
\psfrag{ANA1}[cc][][0.7]{High-SNR analytical SEP}%
\includegraphics [width=3.5in]{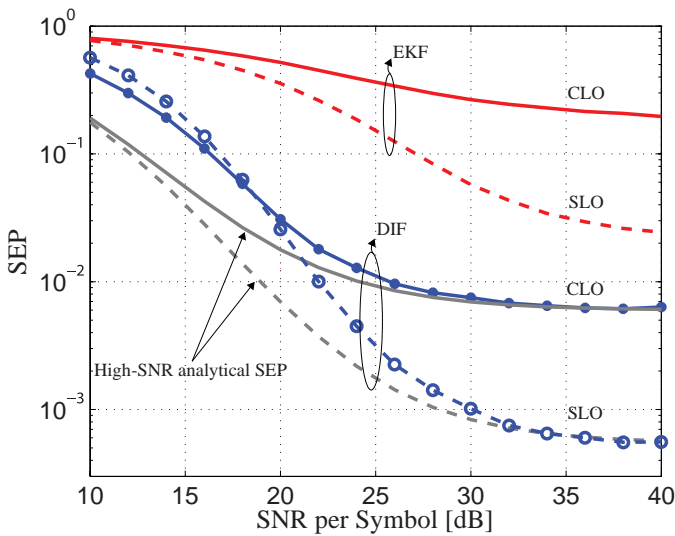}
\caption{SEP versus SNR for CLO and SLO configurations. In this simulation $\sigma^2_\Delta=0.01$~[rad$^2$] and $M=10$.}
\label{fig:EKF_vs_DIFF_vsM}
\end{figure}

Fig.~\ref{fig:EKF_vs_DIFF} compares the SEP of the proposed differential method (DIF), versus the EKF algorithm for a $1\times 10$ SIMO phase-noise channel with CLO configuration for various SNR and $\sigma_\Delta^2$ values. For the EKF algorithm, a pilot symbol is transmitted every $50$ data symbols resulting in a pilot density of $2\%$, while for the differential method no pilot symbols are transmitted. When the phase noise innovation variance is high, e.g., $\sigma_\Delta^2=0.01$~[rad$^2$] or $\sigma_\Delta^2=0.005$~[rad$^2$], the differential method performs significantly better than the EKF algorithm. The EKF method performs close to the differential method for $\sigma_\Delta^2=0.001$ for moderate and high values of SNR. However, as the EKF is highly sensitive to the errors in data detection (i.e., when the transmitted symbol is not a pilot symbol), at low SNR, the performance of the EKF is much worse than the differential method even for low values of~$\sigma_\Delta^2$.

Fig.~\ref{fig:EKF_vs_DIFF_vsM} illustrates the SEP versus SNR of the proposed algorithms for the CLO and the SLO configurations when~$M=10$. It can be seen that the EKF performs significantly better in the SLO configuration compared to the CLO setup. Similar results can be observed for the differential method at high SNR. This gain can be explained as follows: in the SLO case, we receive~$M$ independent noisy observations of the phase of the transmitted signal, which enables us to perform an averaging before detection. As can be seen in~\eqref{eq:pdf_phi_suff_stat_variance}, this manifests as a reduction of the effective variance of the noise that distorts the phase of the signal. Similar observations have been reported in the capacity calculations for the SIMO phase noise channel~\cite{khanzadi2014_SLO_CLO}. However at low SNR, the proposed differential method performs worse in the SLO configuration. This is due to the fact that SNR per antenna becomes quite low as it is scaled by a factor of $M=10$, which renders the approximation in~\eqref{eq:y_phase_approx} to be erroneous. The high-SNR approximation of the SEP performance of the differential signaling method is also presented in Fig.~\ref{fig:EKF_vs_DIFF_vsM}. It can be seen that the analytical SEP matches the simulation results for SNR values around~$30$~[dB] and above. 
\begin{figure}[t]
\centering
\psfrag{serl}[cc][][1]{SEP}%
\psfrag{ml}[cc][][1]{Number of antennas~$M$}%
\psfrag{DIF1}[l][][0.85]{$\sigma^2_{\Delta^{[r]}}=0.01,\sigma^2_{\Delta^{[t]}}=0.01$}%
\psfrag{DIF2}[l][][0.85]{$\sigma^2_{\Delta^{[r]}}=0.01,\sigma^2_{\Delta^{[t]}}=0.008$}%
\psfrag{DIF3}[l][][0.85]{$\sigma^2_{\Delta^{[r]}}=0.01,\sigma^2_{\Delta^{[t]}}=0.006$}%
\psfrag{EF}[cc][][0.85]{Error floor:~$M\to \infty$}%
\includegraphics [width=3.5in]{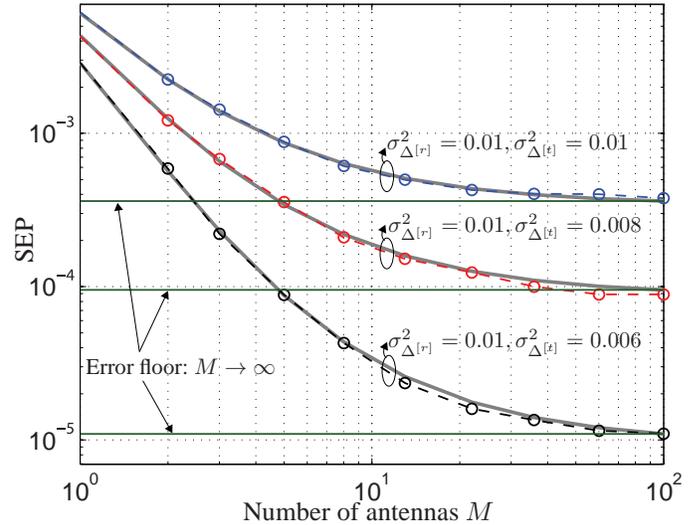}
\caption{Analytical and simulated SEP versus number of antennas and various phase noise innovation variances. In this simulation SNR~=~$40$~[dB].}
\label{fig:DIFF_vsM_40dB}
\end{figure}

Fig.~\ref{fig:DIFF_vsM_40dB} compares the analytical and the simulated SEP of the SIMO system with SLO configuration, for various values of phase noise innovation variance at the transmitter, versus the number of antennas. An error floor is observed by increasing~$M$. We see that the error floor is a function of the phase noise at the transmitter. 
\section{conclusions}
\label{sec:conclusions}
In this paper, we developed a receiver algorithm for a SIMO phase-noise channel that consists of a differential signaling method combined with a two-stage detection of the transmitted signals. This method can be used in strong phase noise scenarios, and when the pilot density mandated by the application is low. It can also be useful in massive-antenna systems because of its low complexity compared to conventional phase noise tracking algorithms like the EKF, and also in the systems whose performance is limited by pilot contamination. The performance of the developed receiver improves by increasing the number of receive antennas due to noise averaging effects. Specially when independent oscillators are used at the receiver, by increasing the number of receive antennas, the effect of phase noise from receiver oscillators averages out. In this configuration, further increasing the number of antennas results in an error floor in the performance, which is a function of the quality of the oscillator used at the transmitter.
\bibliographystyle{IEEEtran}
\bibliography{references}
\end{document}